# On the polyamorphism of fullerite-based orientational glasses

A.N. Aleksandrovskii[1], A.S Bakai[2], D. Cassidy[3], A.V. Dolbin[1], V.B. Esel'son[1], G.E. Gadd[3], V.G. Gavrilko[1], V.G. Manzhelii[1], S. Moricca[3], B. Sundqvist[4].

[1] Institute for Low Temperature Physics & Engineering NASU, Kharkov 61103, Ukraine

[2] National Science Center "Kharkov Institute of Physics & Technology", Kharkov 61108, Ukraine

[3] Australian Nuclear Science & Technology Organisation, Menai, NSW 2234, Australia

[4] Department of Experimental Physics, Umea University, SE - 901 87 Umea, Sweden

Electronic address: dolbin@ilt.kharkov.ua

## Abstract

The dilatometric investigation in the temperature range of 2-28K shows that a first-order polyamorphous transition occurs in the orientational glasses based on $C_{60}$ doped with $H_2$, $D_2$ and Xe. A polyamorphous transition was also detected in $C_{60}$ doped with Kr and He. It is observed that the hysteresis of thermal expansion caused by the polyamorphous transition (and, hence, the transition temperature) is essentially dependent on the type of doping gas.

Both positive and negative contributions to the thermal expansion were observed in the low temperature phase of the glasses. The relaxation time of the negative contribution occurs to be much longer than that of the positive contribution. The positive contribution is found to be due to phonon and libron modes, whilst the negative contribution is attributed to tunneling states of the $C_{60}$ molecules.

The characteristic time of the phase transformation from the low-$T$ phase to the high-$T$ phase has been found for the $C_{60}$-$H_2$ system at 12K.

A theoretical model is proposed to interpret these observed phenomena. The theoretical model proposed, includes a consideration of the nature of polyamorphism in glasses, as well as the thermodynamics and kinetics of the transition. A model of non-interacting tunneling states is used to explain the negative contribution to the thermal expansion. The experimental data obtained is considered within the framework of the theoretical model. From the theoretical model the order of magnitude of the polyamorphous transition temperature has been estimated. It is found that the late stage of the polyamorphous transformation is described well by the Kolmogorov law with an exponent of $n=1$. At this stage of the transformation, the two-dimensional phase boundary moves along the normal, and the nucleation is not important.

## Introduction

In fullerite $C_{60}$, the molecules form a face-centered cubic (*fcc*) lattice. Since the $C_{60}$ molecule has a fifth-order axis of rotation (*C5*), there is no long-range orientational order (LRO) in fullerite. Below the temperature of the glass transition ($T < T_g \approx 90K$) fullerite is an orientational glass. At present, the short- and intermediate range orientational ordering in glasses is not investigated in detail. A two-level model [1,2] is useful to estimate fractions of the "pentagonal" and "hexagonal" mutual orientations of two molecules un the orientational liquid state. In this model cooperative interactions are ignored. Besides, it was speculated theoretically [3,4] and found experimentally [5,6] that the orientational structure of a two-dimensional layer of $C_{60}$ molecules consists of domains (clusters) with narrow boundaries. Conceivably, the formation of $C_{60}$ clusters with certain short - or intermediate - range orientational ordering is possible in the three- dimensional case as well. The short-range orientational order (SRO) is broken at the cluster boundaries. For comparison, in metal glasses with topological or structural (rather than orientational) disorder, the polycluster structure and cluster boundaries were detected by the methods of field-emission microscopy [7-9].

The absence of LRO and orientational structure frustrations lead to the formation of two- or many- level tunnelling states (TS) in glassy fullerites. The distribution of the levels and the characteristic tunneling times is presumably wide [10], but our main interest here is with the states which at low temperatures can make an appreciable or even dominant contribution (as compared to the phonon one) to the thermodynamic coefficients such as the heat capacity and thermal expansion coefficient. Owing to thermally nonactivated tunnel transitions, the TS system attains the thermodynamic equilibrium within quite short times, no matter how low the temperature is. In this case the temperature coefficients are mainly contributed to by the TS's in which the spacings between the lowest energy levels are comparable with the thermal energy and the tunnelling time does not exceed the time of the experiment.

The potential relief of the cluster boundaries in $C_{60}$ has been considered within a two-dimensional model [3,4]. It was found that in the 2D case, the double well states, were separated by low rotational potential barriers with the characteristic heights about $10^2$ times lower at the boundary than in the bulk cluster. This indicates that within the orientational polycluster there exist low-energy TS's at the cluster boundaries, which can generate significant low-temperature effects.

Several low temperature anomalies were detected while investigating the thermal expansion coefficients of pure $C_{60}$ and $C_{60}$ doped with inert gases (He, Ne, Ar, Kr) [11-15]. The most important of them are as follows:

(i). The coefficient of thermal expansion $\alpha(T)$ of pure $C_{60}$ and $C_{60}$ doped with Ne, Ar, Kr is no monotonic dependent on temperature and becomes negative in the finite interval at helium temperatures.

(ii). A hysteresis of $\alpha(T)$ is observed in fullerites doped with He and Kr.

The investigation of the time dependence of the thermal expansion at the jump-like changing temperature, shows that a negative contribution to $\alpha(T)$ appears even when the coefficient itself remain positive.

Taking into account that (i) the TS contribution to $\alpha(T)$ is negative [11-15], (ii) the positive lattice (phonon and libron) contribution $\alpha_L$ decreases rapidly at low temperatures $T << \theta_D, \theta_E$ (where Debye temperature $\theta_D$ and effective Einstein temperature $\theta_E$ are accordingly 54K and 40K [15]) and, (iii) the negative $\alpha(T)$ is caused by the dominant TS contribution, we can obtain at $|\alpha_{TS}| > |\alpha_L|$

$$\alpha(T) = \alpha_{TS}(T) + \alpha_L(T) < 0 \qquad (1)$$

where $\alpha_{TS}$ is the contribution of the TS system to the coefficient of thermal expansion.

The hysteresis of $\alpha(T)$ suggests [15] the existence of at least two phases in the orientational He-$C_{60}$ and Kr-$C_{60}$ glasses, and that the phase transition is of the first order [15]. A transformation of phases which leaves the composition of glasses unaltered is conventionally called polyamorphous. The term "polyamorphism" was first used in [16] and subsequently in [17] to describe the polyamorphous transformation in the amorphous state of the substance. The phenomenon has also been observed in certain liquids and structural glasses [18,19]. It was found also by numerical simulation [20] that amorphous carbon $\alpha$- C exists in two amorphous phases, graphite-like and diamond-like $\alpha - C$. It was revealed that at ambient temperature and pressure changes, the graphite-like phase transforms to the diamond-like one. Fullerite doped with inert gases is the first substance among orientational glasses in which a polyamorphous transition has been detected. It is important to emphasise the diffusion-free kinetics of such transformations. Investigation of the thermodynamics and kinetics of polyamorphous transformations in glasses runs into certain problems which are inherent due the nature of glasses. Since the glass is a system with broken ergodicity, the Gibbs's phase determination and, hence, the methods of conventional statistical physics and equilibrium thermodynamics do not hold. Besides, the state of a nonergodic system is essentially dependent on the thermal history, and the measurement results can greatly be influenced by the duration of the experiment. For this reason the observed polyamorphous transformations are attributed to the rapidly (in comparison with the time of observation) relaxing and interconverting quasi-equilibrium states of the system.

This paper reports experimental and theoretical results on low temperature anomalies in the thermal expansion and polyamorphism of $C_{60}$-based orientational glasses. The objects of dilatometric investigations were solutions of Xe, $H_2$ and $D_2$ in $C_{60}$. This is the first inquiry into the thermal expansion of these solutions.

We propose a theory that includes both a general approach to polyamorphous transformations and a consideration of thermal expansion and phase transformation kinetics (Section 2), which allows us to analyze the experimental data (Section 3). Comments and conclusions complete the paper.

## 1. Experimental technique and results.

The linear thermal expansion coefficient $\alpha(T)$ was investigated using a high-sensitivity capacitance dilatometer [21] and the technique described in [11]. Since the used pure $C_{60}$ and gas-saturated $C_{60}$ have a cubic lattice, their thermal expansion is isotropic and is characterised by a scalar $\alpha(T)$.

### 1.1 Xe − $C_{60}$ system.

The $C_{60}$ powder (99.99%) with an average grain size of about 100 μm (SES Co., USA) was intercalated with xenon and then compacted. The intercalation was performed for 36 hours at a Xe pressure of ~200 MPa and a temperature of 575°C. According to the thermal gravimetric analysis (TGA), about 30% of the octahedral cavities of $C_{60}$ were filled with Xe, and in agreement with previous studies [22].

Since the process of Xe-$C_{60}$ dissolution produces considerable deformation of the $C_{60}$ lattice [22], extreme care was taken to prepare the Xe-$C_{60}$ samples for dilatometric investigation.

In our previous studies [11,13-15] polycrystalline specimens were formed by pressing fullerene powder in a cylindrical die for 30-45 minutes at an effective pressure between 0.5 and 1 GPa. The die consisted of an inner ring with a cylindrical bore and a conical outer surface, fit into an outer cylinder hardened steel which provided support for the inner pressure. The inner ring was split into four sections. After pressing, the conical inner part was carefully forced out of the outer ring, after which the four sections of the inner ring could be removed with minimum damage to the specimen. The piston used was also made from hardened steel.

The present compound was rather difficult to press into cylinders than those previously studied. Pure $C_{60}$ is a soft solid which easily deforms by plastic flow, and many high-pressure studies have been carried out using the material itself as a pressure transmitting medium [23]. Thin solid polycrystalline plates are easily produced by applying a nominal (force-over-area) pressure of up to 1 GPa to $C_{60}$ in the powder form, which then deforms much like a hard wax. However, because $C_{60}$ hardens appreciably when deformed, in order to produce homogeneous cylindrical specimens with a height approximately equal to the diameter, as used in our measurements of the thermal expansion, it was necessary to fill the cylinder gradually. In all experiments we have filled the cylinder in ten or more steps, packing the powder well before the next batch was poured in. Even then, when using this method, it has been necessary to handle the specimen carefully when removing it from the die, since rough handling usually resulted in the breaking of the specimen. The most common form of fracture observed were cracks perpendicular to the axis, but sometimes conical fractures also occurred, producing low cones with the end surfaces as base.

These difficulties were much more pronounced in the present experiments with the Xe doped material. The first attempt to form a cylindrical specimen failed; after pressing, the specimen broke into several pieces. After crushing the material, a second attempt was made, this time successfully. The cylindrical sample was then sent from Sweden to Ukraine for study, but on arrival it had broken into several pieces and was subsequently returned to Umeå. A third, very careful attempt was then made, with a newly cleaned die to minimize friction and again with great care in handling. This resulted in a final specimen which was allowed to rest for more than 24 h to see if spontaneous cracks would appear. Since no cracks were observed, the specimen was placed between two teflon cylinders and wound with thin Teflon film until it fitted perfectly in a glass tube. Cotton wool was added to give a slight pressure on the specimen during transport to ILTPE (Ukraine).

Despite the precautions, the specimen mailed to Ukraine broke into two unequal pieces nearly in parallel to the base. The larger piece was used for thermal expansion investigation. The final sample was a cylinder 5 mm high and 10 mm in diameter. Before measuring, it was kept in a Xe atmosphere (760 torr) in a glass ampoule at room temperature. The sample was then transferred to the measuring cell of dilatometer. The procedure was performed in the air and took 20 minutes. The cell with the sample was then successively evacuated, filled with Xe at 760 torr and sealed. The sealed measuring cell with the sample in the Xe atmosphere was cooled to 160K. At this temperature, the measuring cell was evacuated again and cooled down to liquid helium temperature.

The vacuum in the cell was maintained at $1 \cdot 10^{-5}$ torr during the whole experiment. The cooling from room temperature to liquid helium temperature took 12 hours.

The temperature dependence of the $\alpha(T)$ measured on the Xe-$C_{60}$ sample in the range 2-28K is shown in Fig.1 (the broken arrows point to the direction of the temperature change during the experiment). It is seen that on heating and subsequent cooling of the sample, the $\alpha(T)$ has a considerable hysteresis. As mentioned above, the hysteresis of the temperature dependence $\alpha(T)$ of Xe-doped $C_{60}$ indicates the existence of two phases of the orientational glass. For comparison, the same figure illustrates the thermal expansion of a pure $C_{60}$ sample (broken curve 3) compacted from the pure $C_{60}$ powder that was used for preparing (Xe)$_{0.3}$-$C_{60}$, by an identical procedure. In pure $C_{60}$ the behaviour of the $\alpha(T)$ is similar on heating and cooling. In both cases, the step-like change in the sample temperature $\Delta T$, was kept approximately the same. On increasing temperature, the step changed from 0.3K to 1.5K.

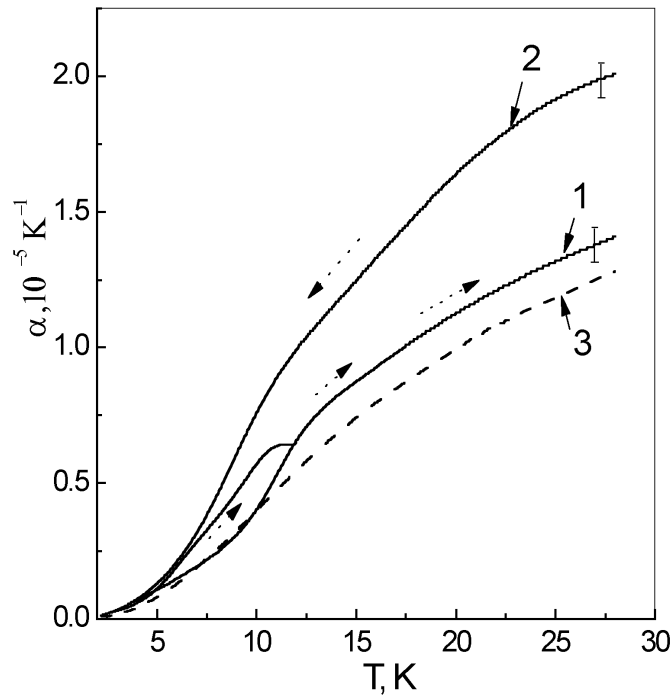

Fig.1. Temperature dependence of the linear thermal expansion coefficient of (Xe)$_{0.3}$-$C_{60}$ and pure $C_{60}$:

1 – heating of (Xe)$_{0.3}$-$C_{60}$ ;

2 – cooling of (Xe)$_{0.3}$-$C_{60}$ ;

3 – pure $C_{60}$.

At 5 and 12K, curve 1 (Fig.1) is branching. The upper branch appeared with rising temperature immediately after the first cooling of the sample from room temperature to 4.2K. The lower branch emerged after repeated cooling to 2.2K and subsequent heating. It is evident that the hysteresis observed in curves 1 and 2 is determined by the temperature prehistory of the sample.

It is seen in Fig.1 that the thermal expansion of (Xe)$_{0.3}$-$C_{60}$ is always positive over the range of temperatures studied. However, on heating the sample by $\Delta T$, the time dependence of the thermal expansion exhibits two processes (see Fig.2) with different characteristic times. The contribution of the faster process to the thermal expansion was positive whilst that of the slower process was negative. In [15] the positive contribution has been attributed to low-frequency excitations (phonons and librons), while the negative contribution is related to tunnel reorientations of the $C_{60}$ molecules.

The positive contribution to the thermal expansion of the (Xe)$_{0.3}$-$C_{60}$ sample exceeded the negative one over the whole temperature range of the experiment. As a result, the total thermal expansion coefficient is always positive.

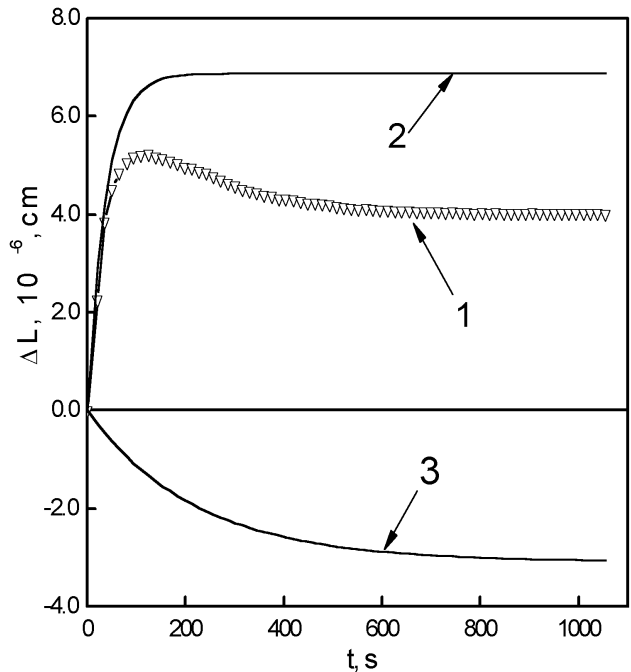

Fig.2. Characteristic time dependence of the sample length on heating
(Xe)$_{0.3}$-C$_{60}$ by $\Delta T$:
1 – experimental results;
2 – positive contribution to thermal expansion;
3 – negative contribution to thermal expansion.

On cooling the (Xe)$_{0.3}$-C$_{60}$ sample there is only a positive contribution to the thermal expansion.

At constant temperature $\alpha(t)$ dependence can be described by :

$$\alpha(t) = \frac{1}{\Delta T} \cdot \frac{\Delta L}{L} = A(1\text{-}exp(\text{-}t/\tau_1)) + B(exp(\text{-}t/\tau_2)\text{-} 1) , \qquad (2)$$

where the first and second terms on the equation's right-hand side describe the positive and negative contributions respectively; $A$ and $B$ are the absolute values of the corresponding contributions at $t \to \infty$; and $\tau_1$ and $\tau_2$ are the characteristic relaxation times for these contributions. The $B/A$ value is the ratio of the negative to positive contributions to the $\alpha(T)$.

Using the data processing procedure of [15], we can evaluate the characteristic times of the processes responsible for the thermal expansion of the (Xe)$_{0.3}$-C$_{60}$ sample and evaluate the positive and negative contributions as a function of temperature. The characteristic times of the positive and negative contributions, $\tau_1$ and $\tau_2$, are shown in Fig.3. For comparison, the figure includes the characteristic times of the positive and negative contributions to the thermal expansion of the (Kr)$_{0.625}$-C$_{60}$ sample [15]. It is seen that in interval 12-22K the characteristic times of the negative contribution to the thermal expansion of the (Xe)$_{0.3}$-C$_{60}$ sample are smaller than those for the (Kr)$_{0.625}$-C$_{60}$ sample. The characteristic times of the positive contributions in these samples coincide within the measurement error.

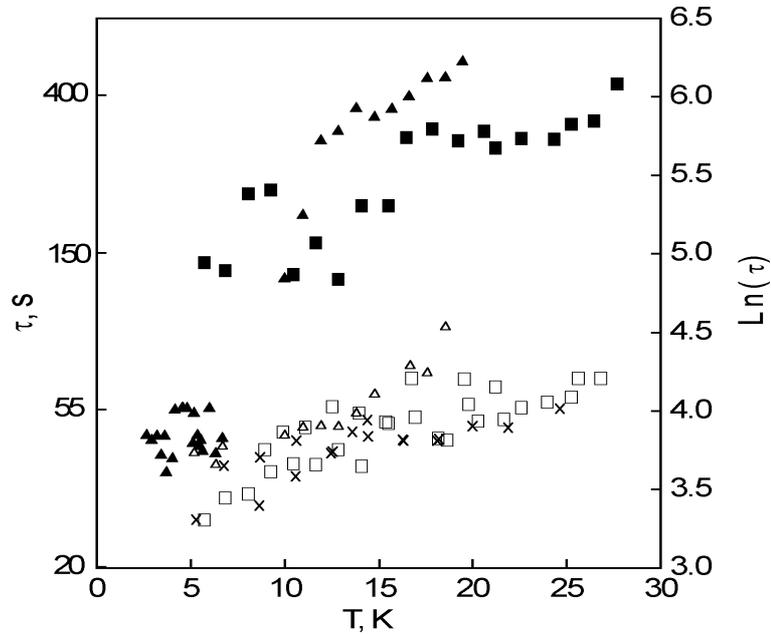

Fig.3. Characteristic times of positive and negative contributions to thermal expansion of $C_{60}$ samples intercalated with xenon and krypton:

$\Delta$, $\times$ - characteristic times of positive contributions $(Kr)_{0.625}$-$C_{60}$
($\Delta$ - on heating, $\times$ - on cooling),

$\square$ - characteristic times of positive contributions $(Xe)_{0.3}$-$C_{60}$,

$\blacktriangle$ - characteristic times of negative contributions $(Kr)_{0.625}$-$C_{60}$,

$\blacksquare$ - characteristic times of negative contributions $(Xe)_{0.3}$-$C_{60}$.

The temperature dependence of the ratio $B/A$ of the $(Xe)_{0.3}$-$C_{60}$ and $(Kr)_{0.625}$-$C_{60}$ [15] samples is shown in Fig.4. For $(Xe)_{0.3}$-$C_{60}$ the highest value of the ratio is considerably smaller in comparison with what is observed for $(Kr)_{0.625}$-$C_{60}$, and is shifted towards higher temperatures. We believe that the reason may be as follows. The gas-kinetic diameter of Xe atoms is larger than that of Kr atoms. When impurity atoms penetrate into the octahedral cavities of fullerite, the distance between the $C_{60}$ molecules surrounding the impurity atom increases, and for the case of fullerite saturation with Xe atoms, results in a larger spacing than in the Kr case. Correspondingly, for Xe-$C_{60}$ solutions, the rotational barrier for the $C_{60}$ molecules is lower. As a result, the tunnel splitting of the energy levels of $C_{60}$ rotation increases and the manifestation of this is that the negative thermal expansion effect shifts towards higher temperatures (see Section 2).

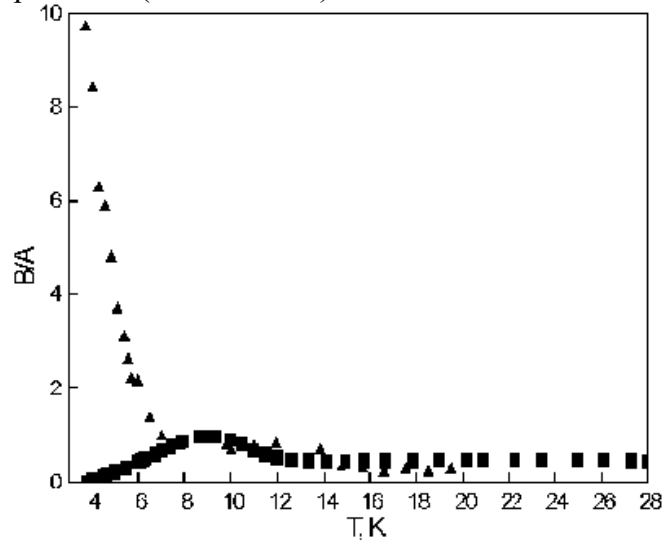

Fig.4. The absolute value of the ratio between the negative and positive contributions to the thermal expansion of $(Xe)_{0.3}$-$C_{60}$ ($\blacksquare$) and $(Kr)_{0.625}$-$C_{60}$ ($\blacktriangle$) solutions. The coefficient of thermal expansion becomes negative when $B/A > 1$.

Our investigations on the $(Xe)_{0.3}$-$C_{60}$ (this study) and $(Kr)_{0.625}$-$C_{60}$, He-$C_{60}$, Ne-$C_{60}$, Ar-$C_{60}$ [13-15] systems lead to the following conclusions. The positive (phonon and libron) contribution to the thermal expansion of the above solutions comparatively weakly depends on the type and concentration of the inert gas dissolved in fullerite. These factors, however, influence significantly the negative contribution caused by tunnel reorientation of the $C_{60}$ molecules and the $\alpha(T)$ hysteresis caused by phase transformations of the orientational glasses.

### 1.2  $D_2$-$C_{60}$ system.

The initial sample of pure $C_{60}$ was compacted from $C_{60}$ powder (Term, USA, Berkeley, CA). Before saturation with $D_2$, the sample was dynamically evacuated ($1\cdot10^{-3}$ torr, T=250°C) for 48 hours to remove the gas impurities. Thereupon the linear thermal expansion coefficient of pure $C_{60}$ was measured. It was positive in the whole temperature interval (2.2-24K). No hysteresis in the temperature dependence of the thermal expansion coefficient was observed on heating or cooling the sample. The negative contribution to the thermal expansion was also absent in the whole $T$-range. When these measurements were completed, the measuring cell with the sample was filled with $D_2$ at room temperature up to 760 torr. Under this condition, the sample was saturated with $D_2$ for 15 days. Then the measuring cell with the sample was cooled slowly down to 25K for 8 hours.

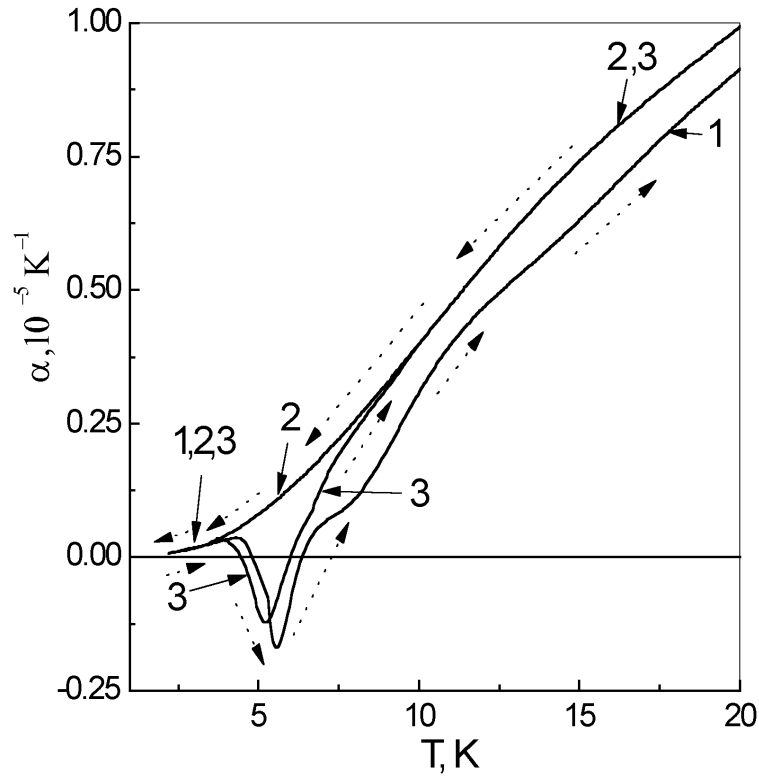

Fig.5. Temperature dependence of linear thermal expansion coefficient of $D_2$-doped $C_{60}$:
1 – heating $D_2$-$C_{60}$;
2 – cooling $D_2$-$C_{60}$;
3 – heating $D_2$-$C_{60}$ after five-months $D_2$ – desaturation at room temperature.
Broken arrows show directions of temperature variations in experiment.

On reaching $T$=25K, the cell with the sample was evacuated to no less than $1\cdot10^{-5}$ torr and the cooling was continued down to $T$=2.2K, at which the sample was kept for 5 hours before the dilatometric measurement.

The temperature dependence of $\alpha(T)$ of $D_2$–saturated $C_{60}$ is shown in Fig.5. It is seen that on heating (curve 1) and subsequent cooling (curve 2) the $D_2$-$C_{60}$ sample there was a hysteresis in the $\alpha(T)$ above 3.5K (an indication of phase transformations in the orientational glass).

On heating $D_2$-$C_{60}$, both negative and positive contributions to the expansion were detected. The temperature dependence of the ratio $B/A$ of the $D_2$-$C_{60}$ sample is shown in Fig.6. A sharp $B/A$ maximum is observed near 5K. On heating $D_2$-$C_{60}$, the absolute value of the negative contribution is higher in the interval 4.5-6.3K.

As was stated earlier [15], the hysteresis in the temperature dependence of thermal expansion, and hence, orientational polyamorphism, appear when the size of the introduced particle exceeds that of the interstitial cavity. In this case, the hysteresis even appears when comparatively small $D_2$ molecules are introduced into the fullerite lattice. This may indicate that unlike Ne, Ar, Kr and Xe atoms [22], $D_2$ molecules occupy not only the octahedral cavities of $C_{60}$, but also the much smaller tetrahedral ones as well.

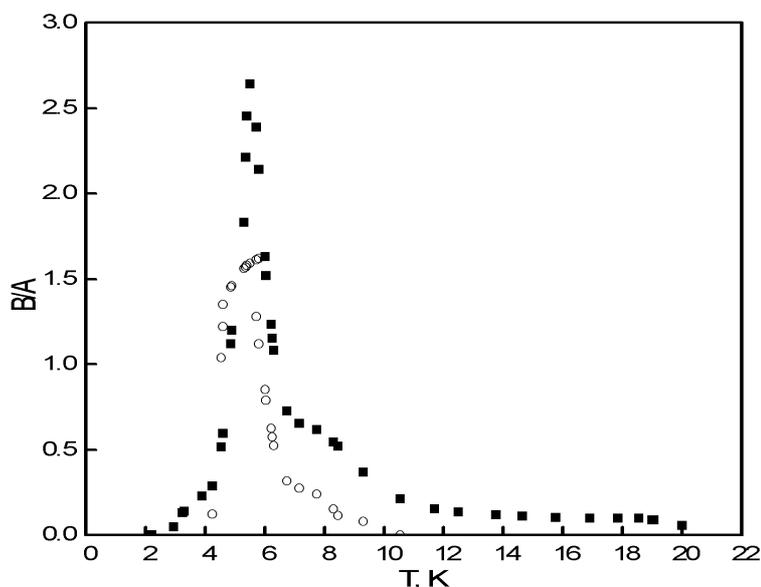

Fig.6. Absolute values of the ratio of negative and positive contributions to the thermal expansion of $D_2$-$C_{60}$ (■ – before desaturation, ○ – after $D_2$ desaturation for 5 months at room temperature). The thermal expansion coefficient becomes negative when $B/A$>1.

Note that earlier in [15], we also detected a hysteresis in the temperature dependence of the thermal expansion of the He-$C_{60}$ solution. The assumption that He atoms occupy both octahedral and tetrahedral interstitial cavities in $C_{60}$ is substantiated in [25].

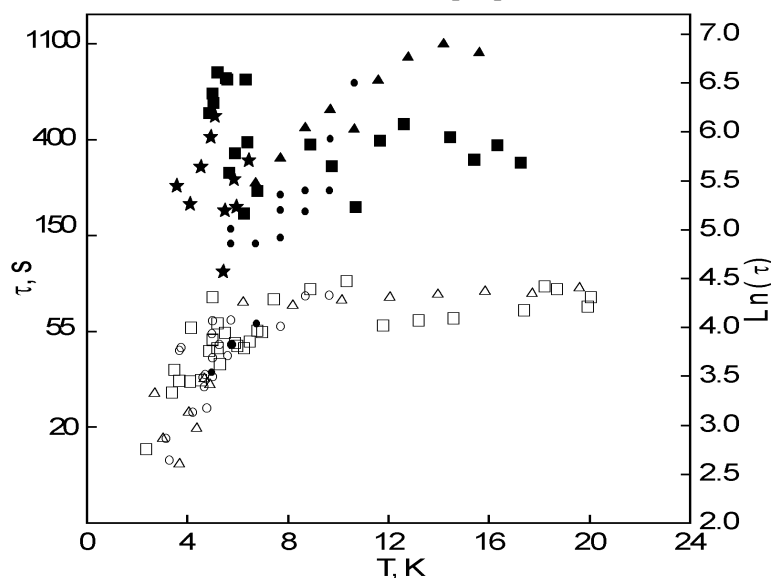

Fig. 7. Characteristic times of positive and negative contributions to the thermal expansion of $C_{60}$ doped with $D_2$ and He:

Δ, ○ – characteristic times of positive contribution of He-$C_{60}$ [15]

(Δ - before partial desaturation, ○ - after partial desaturation);

□ – characteristic times of positive contribution of $D_2$-$C_{60}$;
▲ – characteristic times of negative contribution of He-$C_{60}$ [15];
■ - characteristic times of negative contribution of $D_2$-$C_{60}$;
★- characteristic times of negative contribution of $D_2$-$C_{60}$ after 5-months' desaturation at room temperature.

The temperature dependences of the characteristic times of the negative and positive contributions, $A$ and $B$, of the $D_2$-$C_{60}$ sample are shown in Fig. 7. It is seen that the largest characteristic times of the negative contribution are observed at 5K.

Recall the sharp maximum in the temperature dependence of the ratio $B/A$ just at this temperature (see Fig. 6).

When the dilatometric investigation was completed, the $D_2$-$C_{60}$ sample was partly desaturated. To do this, a vacuum no worse than $10^{-1}$ torr was maintained in the dilatometric cell with the sample for 150 days at room temperature. Under these conditions the desaturation temperature also corresponded to the temperature at which the sample was saturated with $D_2$. The desaturation took an order of magnitude more time than the exposure in the $D_2$.

A further dilatometric investigation showed that the removal of $D_2$ was not complete even after the above procedure. The hysteresis re-appeared in the temperature dependence of the thermal expansion coefficient (Fig. 5, curve 3). But now the hysteresis and the temperature region of its existence were considerably smaller. In contrast to the $D_2$-saturated sample, in which the negative contribution to the thermal expansion persisted over the whole investigated $T$ –range investigated, the negative component of the thermal expansion of the partly desaturated sample was observed only in the temperature interval of hysteresis 3.5-9K, and its absolute values were lower than in the case of $D_2$-$C_{60}$ (See Fig. 6). The positive contribution remained unaltered within the experimental error.

At room temperature and atmospheric pressure, fullerite does not form chemical bonds to hydrogen molecules [24]. The incomplete removal of $D_2$ from $C_{60}$ during a prolonged desaturation at room temperature supports the aforesaid assumption that $D_2$ penetrates into the tetrahedral cavities of $C_{60}$. We can expect that the energy of the interaction between the $D_2$ and $C_{60}$ molecules is higher in the comparatively small tetrahedral cavities than in the octahedral ones. The $D_2$ molecules therefore have a lower chance to leave the tetrahedral cavities.

Our assumption that $D_2$ occupies the tetrahedral cavities of $C_{60}$ is in conflict with the conclusions in [24] where neutron scattering investigation detected hydrogen and deuterium only in the octahedral cavities of $C_{60}$. But the gas saturation of $C_{60}$ powder in [24] only lasted for 5 hours, which might be insufficient to saturate the tetrahedral interstitial cavities of the subsystem with hydrogen and deuterium. Note that in [25] intercalation of $C_{60}$ powder with helium at normal temperature and pressure lasted for 4000 hours. It was concluded that the time taken to saturate the tetrahedral subsystem of interstitial sites with helium was two orders of magnitude longer than that needed to saturate the octahedral subsystem.

### 1.3 $H_2$-$C_{60}$ system

After $D_2$ desaturation at 250ºC, the resultant $C_{60}$ sample was used to investigate the thermal expansion of a $H_2$-$C_{60}$ solution. The $C_{60}$ sample was saturated directly with $H_2$, in the measuring cell held at 20ºC and a pressure of 760 torr, for 13 days. The sample was then cooled down to liquid helium temperature using the procedure described in the previous section. The thermal expansion of $H_2$-$C_{60}$ was investigated in the temperature interval of 2.2-22K. The results are shown in Fig. 8.

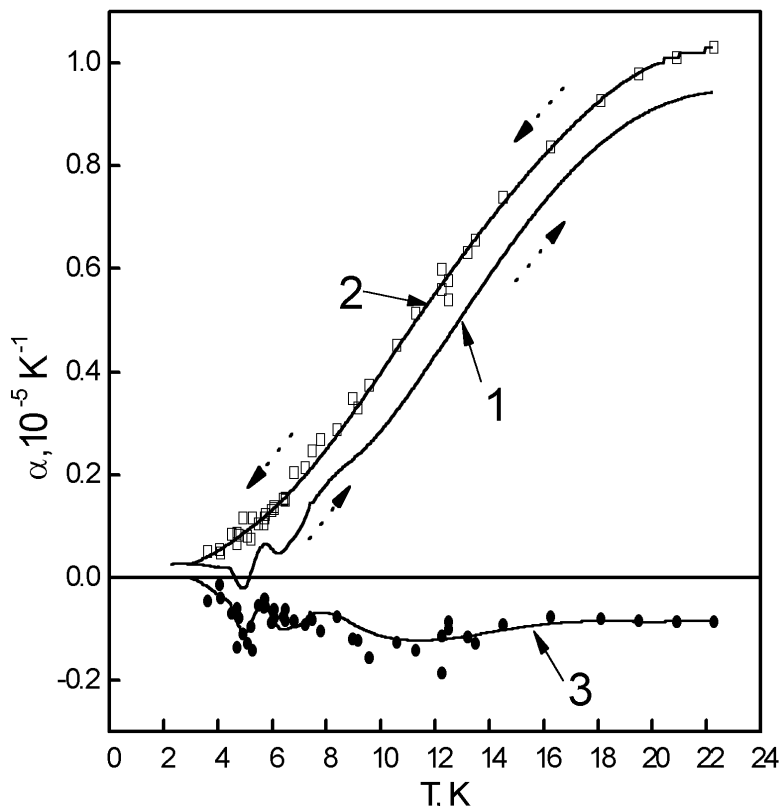

Fig.8. Temperature dependence of the linear thermal expansion coefficient of
$H_2$-$C_{60}$:
curve 1 – data taken on heating the sample;
curve 2 – data taken on cooling the sample;
curve 3 – the difference between curves 1 and 2;
□ – positive contribution to the thermal expansion of $H_2$-$C_{60}$,
● – negative contribution to the thermal expansion of $H_2$-$C_{60}$.

It is seen that the hysteresis in the temperature dependence of the $\alpha(T)$ of $H_2$-$C_{60}$ persists over the whole temperature interval. As in the case of $D_2$-$C_{60}$, on heating $H_2$-$C_{60}$, both negative and positive contributions with different characteristic times are present in the thermal expansion. On cooling, however, only a positive contribution is observed. Curves 1 and 2 in Fig. 8 were obtained by a polynomial approximation of experimental $\alpha(T)$ values, using a least squares fit technique. Curve 3 is the difference between curves 1 and 2. The positive (squares) and negative (circles) contributions to the thermal expansion with heating the sample, were evaluated through processing the time dependences of the length variations of the $H_2$-$C_{60}$ sample, using Eq. (1). The heating-induced positive contributions to the thermal expansion of $H_2$-$C_{60}$ (open squares), and the $\alpha(T)$ measured on cooling agree within the experimental error over the whole temperature interval. This supports our assumption that the phonon and libron contributions to the $\alpha(T)$ are insensitive to the type of orientational glass. Besides, the negative contribution measured on heating (filled circles) of $H_2$-$C_{60}$ agrees, within the experimental error, with curve 3 in Fig.8, which is the difference between curves 1 and 2. This means that the negative contribution is precisely responsible for the decrease in the $\alpha(T)$ obtained on heating the sample (Fig. 8, curve 1 as compared to curve 2).

The $\alpha(T)$'s of $D_2$-$C_{60}$ (solid lines 1,2) and $H_2$-$C_{60}$ (broken curves 3, 4) are shown in Fig. 9. The curves taken on cooling $H_2$-$C_{60}$ and $D_2$-$C_{60}$ agree within the experimental error over the whole temperature range. On heating, the $H_2$-$C_{60}$ and $D_2$-$C_{60}$ samples exhibit quite different behaviour of the $\alpha(T)$ in the interval 5-9K.

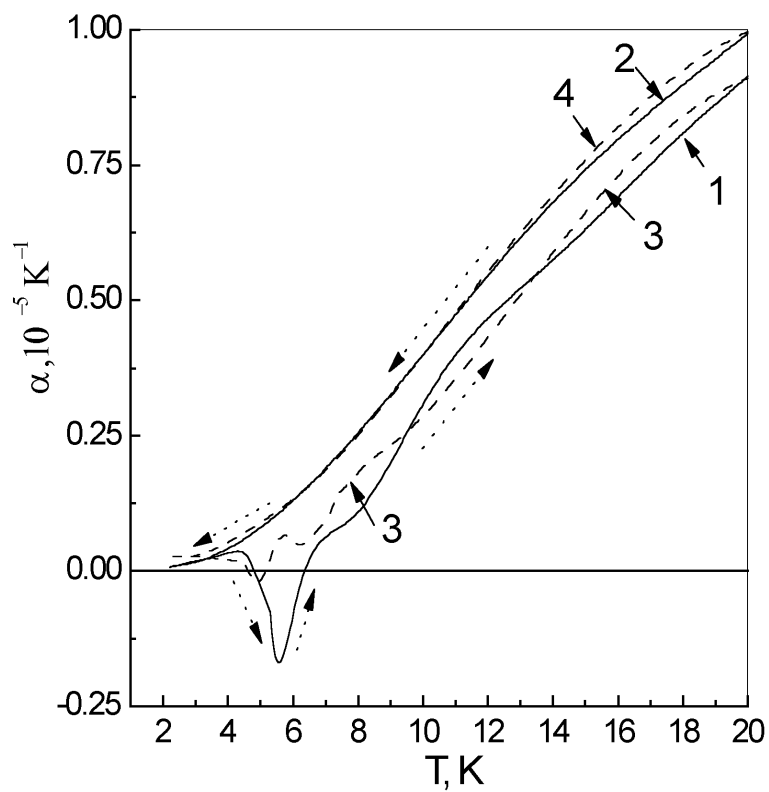

Fig. 9. Temperature dependences of $\alpha(T)$ for $H_2$-$C_{60}$ and $D_2$-$C_{60}$:
1 - heating of $D_2$-$C_{60}$;
2 - cooling of $D_2$-$C_{60}$;
3 - heating of $H_2$-$C_{60}$;
4 - cooling of $H_2$-$C_{60}$.

The temperature dependences of the ratio between the negative and positive contributions to the thermal expansion of the $H_2$-$C_{60}$ sample are shown in Fig. 10.

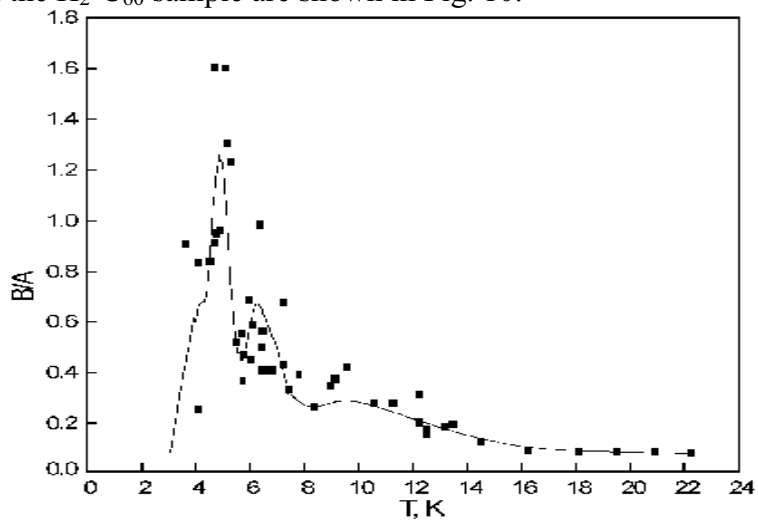

Fig.10. The absolute values of the ratio between the negative and positive contributions to the thermal expansion of $C_{60}$ saturated with $H_2$ (■). The curve in fact corresponds to the ratio of the absolute value of curve 3 to that of curve 2 as were shown previously in Fig. 8.

The characteristic times of the positive and negative contributions to the thermal expansion of $H_2$-$C_{60}$ and $D_2$-$C_{60}$ are shown in Fig.11.

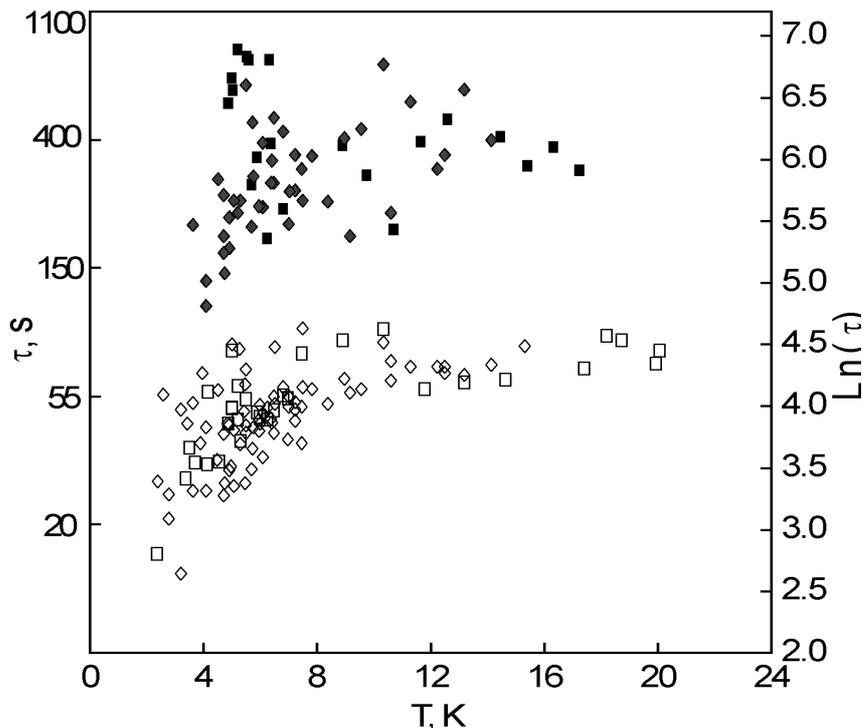

Fig.11. Characteristic times of positive and negative contributions to the thermal expansion of $H_2$-$C_{60}$ and $D_2$-$C_{60}$:

◊- for positive contribution, $H_2$-$C_{60}$,

□ - for positive contribution, $D_2$-$C_{60}$,

♦ - for negative contribution, $H_2$-$C_{60}$,

■ - for negative contribution, $D_2$-$C_{60}$.

They are seen to coincide for both the solutions within the experimental accuracy. The dynamics of the processes responsible for the hysteresis in the temperature dependence of the $\alpha(T)$ of $H_2$-$C_{60}$, was investigated in the following experiment. After keeping the sample at 4.2K for four hours, its temperature was raised to 9.5K during 20 min. The sample was then thermo - cycled in the interval 9.5-13.5K at a step of 2K. Duration of one step was half an hour. The thermo - cycling procedure thus involved two heating steps followed by two cooling steps. In doing so, we obtained the $\alpha(T)$ values at 10.5 and 12.5K. As the temperature increased, the negative contribution decreased with time whilst the positive contribution did not vary. On lowering the temperature, only the positive contribution is present. It was thus possible to extract the time dependence of the normalized negative contribution $B'$ to the thermal expansion of $H_2$-$C_{60}$ at 10.5K and 12.5K and as shown by the triangles and circles, respectively in Fig.12. The normalized $B'$-values are the ratios of the negative contribution at the moment $t$ to the negative contribution at the initial instant of time ($t$=0 is the time corresponding to stabilisation of $T$ = 11.5K).

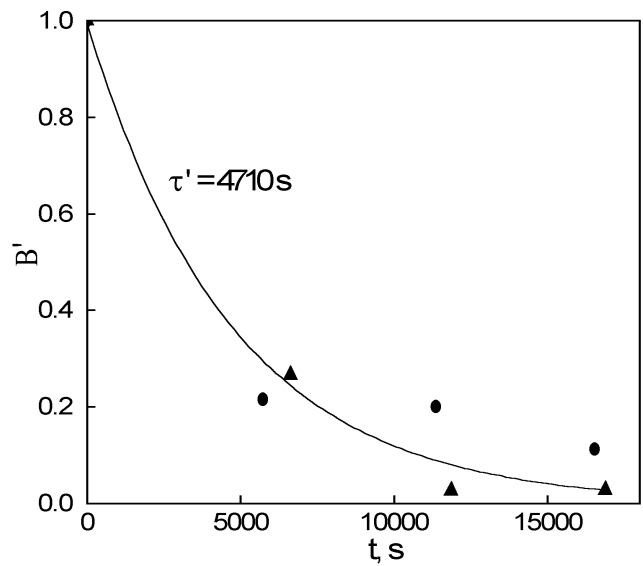

Fig.12. Dependence of normalized negative contributions to the $\alpha(T)$ of $H_2$-$C_{60}$ upon the current time point of the experiment obtained during thermocycling and within the temperature range 9.5-13.5K:
▲ - normalized negative contribution at 10.5K;
● - normalized negative contribution at 12.5K;
The solid curve is the exponential approximation of the results.

The time dependence of the experimentally measured negative contributions to the thermal expansion of $H_2$-$C_{60}$ was approximated by an exponent (see Fig.12). The characteristic time $\tau'$ of this process found by the exponential approximation is 4700s. The time $\tau'$ is actually the characteristic time of the phase transformation in the investigated orientational glass. Note that the characteristic time of reorientation of the $C_{60}$ molecules in the $H_2$-$C_{60}$ system at $T$=11.5K is an order of magnitude shorter ($\approx$400 s, see Fig.11).

## 2. Theory.
## 2.1. Phase transition in systems with broken ergodicity.

### 2.1.1  Specific features of systems with broken ergodicity.

Statistical physics based on the Gibbs microcanonical distribution postulates that any system is ergodic, i.e. its phase trajectory covers the isoenergetic surface, $E = \varepsilon(p,q)$, densely and throughout, and along the phase trajectory, the time average is equal to the average over the isoenergetic surface.

On using the canonical distribution, the averaging over the isoenergetic surface is replaced by the ensemble averaging. In systems with broken ergodicity [26], the isoenergetic surface can be subdivided into simply connected regions (basins), in which, on the one hand, a considered phase trajectory may never go beyond the boundaries of the basin it has been introduced into, during the observation time $\tau_{obs}$ and, although on the other hand, it may still have enough time to cover densely the whole basin.

It is seen that the definition of a basin inherently has to consider the parameter $\tau_{obs}$. Sometimes (e. g., in systems with spontaneous broken symmetry) the barriers separating the basins are infinitely large in the thermodynamic limit. In this case, the phase trajectory would never leave the basin into which it was brought, by the process of evolution. The subdivision of the isoenergetic surface into basins corresponding to various possible states in this case would therefore be independent of $\tau_{obs}$.

Glasses are typical systems with broken ergodicity whose properties depend on the time of observation. This statement itself implies the existence of finite barriers between the basins.

Nevertheless, the properties of aged and annealed (at $T<T_g$) glasses vary very little with time, and such changes can usually be neglected. It can therefore be assumed that the phase trajectory of such a glass system and during the period of observation belongs to one invariable basin. As for the total number of basins possible, or equivalently the number of structural states of the glass available, this turns out to be exponentially large and is given by:

$$W(N) = \exp(\varsigma N) \tag{2}$$

Here $N$ is the number of molecules in the considered system and $\varsigma > 0$. $\varsigma$ is usually of the order of unity. In an ergodic system, the quantity $\varsigma$ would describe the configuration entropy (assuming that the Boltzmann constant is unity). In the glass, where the system is arrested in one of the basins, $\varsigma$ describes the complexity of the structure [26].

The free energy of a glass arrested in the i-th basin, $G_i(P,T)$, can be calculated taking into account the states belonging only to this basin. Since the system has no symmetry, its states, possessing a unique energy, are not equivalent thermodynamically. The numerously found deepest minima of the free energy in the isoenergetic layer, in themselves make up a metabasin of states of which the glass is found invariably in one of these states.

### 2.1.2. Short- and intermediate- range order, isoconfigurational states.

**Short-range order.** The glass structure has no long-range order and is therefore characterized by the short-range and intermediate-range orders. The short-range order (SRO) is determined by the correlation in the mutual arrangement of the neighbouring molecules (mutual orientation). Let us consider the SRO in a certain site of a *fcc* lattice. Because of the short-range interaction of the molecular forces, the highest contribution to the free energy is made by the interactions with the nearest surroundings, i.e. with those molecules forming a coordination polyhedron. The mutual orientation of the molecules within the coordination polyhedron is definitely responsible for the SRO. There are a finite number of non-equivalent orientational configurations of molecules in the coordination polyhedron that correspond to the free energy minima. Hence, there are a finite number of SRO types, which can undergo thermally activated rearrangements. Although the coordination polyhedron is a natural structural element, the molecular associates forming significant many-particle orientational correlations can be inconsistent with the coordination polyhedra. For simplicity, we assume that these orientational associates are equal in size to the coordination polyhedra.

The associativity of molecules in terms of the SRO is also taken into account in the description of the cooperative phenomena in ordinary vitrifying liquids (see [8, 27, 28]). In this case the description of the thermodynamic properties is reasonable if we assume that the average number of molecules in the associate, $k_0$, is about 10, i.e. the associates include the molecules of the first coordination sphere. Assuming that the orientational associates having different types of SRO, are independent statistically, we can estimate their equilibrium concentration in the melt using Gibbs statistics. In the cell approximation, the partition function of the melt $Z(P,T;N,k_o)$ is:-

$$Z(P,T;N,k_0) = [z(P,T;k_0)]^{N/k_0}$$

$$z(P,T;k_0) = \sum_i \exp[-k_0 g_i(P,T)\beta], \ \beta = 1/T \tag{4}$$

Here $z(P,T;k_0)$ is the partition function of a cell consisting of $k_0$ molecules; $g_i(P,T)$ is the mean free energy per molecule in the *i*-th type associate and $N$ is the number of molecules. Boltzmann constant is taken as 1 also in Eq. 4, as well as ongoing in this paper.

According to Eq. (4), the fraction of the *i*-th type of orientational associate out of a considered total number possible of $n_s$, is:

$$c_i(P,T) = \exp[-k_0 g_i(P,T)\beta]z^{-1}(P,T;k_0) \tag{5}$$

The number of the associate types, $n_s$, is a characteristic of substance.

Remember that this expression holds only for the melt (ergodic), where Gibbs statistics (Eq. 4) are valid. If we assume that the loss of ergodicity during the glass transition does not cause significant changes in the SRO, then at a temperature $T < T_g$, the corresponding fraction of the $i$-th type of orientational associate in the glass is given by:

$$c_{i,g} = c_i(P, T_g) \qquad (6)$$

**Intermediate- range order.** In accord with the approximation used the glass structure is considered as consisting of $N/k_0$ cells. Each of the cells possesses one of the possible SRO's correlations of orientational order can exist. Although the orientational pair correlations become weaker with distance, the correlations between the mutual positions of various types of associates can extend to large (as compared to associate sizes) distances. This phenomenon is observed in ordinary glass forming liquids and polymers [28,29]. Since the interaction with the molecules of the neighboring associate is weaker than that inside the associate, the formation of long-range pair correlations has only a slight effect on the free energy and, hence, on the fraction of a particualr type of associate in the glass as determined by Eq. (6). However, these correlations are still important structural characteristics of the glass. The pair correlations of the mutual positions of various types of associates are

$$w_{ik}(r) = <c_{i,g}(x)c_{k,g}(x+r)>, \; r > r_a \approx ak_0^{1/3} \qquad (7)$$

Here $r_a$ is the size of cell (size of associate) specified by a SRO and $a$ is the size of molecule; the local orientational order and the angle brackets indicate spatial averaging. When correlations are absent, then $w_{ik}(r) = \overline{c_{i,g}} \cdot \overline{c_{k,g}}$ ($\overline{c}_{i,g} \equiv <c_{k,g}(x)>$). The first maximum of $w_{ik}(r)$ evidently is at $r_p \approx 2r_a$. The values of pair correlation functions at $r_p$ are therefore important structural characteristics. Let us denote them as $w_{ik}(r_p)$. The number of the neighbouring associates is $k \approx 2^3 \sim 10$. For the $i$-th type associate, within the sphere of radius $r_p$ the fraction $c_{i,g}$ is nearly equal to $w_{ii}(r_p)$ provided that the associate of this type occupies the position at the center of the sphere.

On the intermediate scale, the orientational structure is characterized, along with $w_{ik}(r_p)$, by higher-order correlators.

**Isoconfigurational structures**. The static properties of orientational configurations are determined by the magnitudes $\left\{ c_{i,g}, w_{ik}(r) \right\}$ and by higher-order correlators. Two states of the glass will be considered isoconfigurational in the $m-$order if their first $m-$correlators coincide. Differences between correlators higher than $m$, reflect the distinctions between the structures and properties of the two states of the glass. On the other hand two isoconfigurational structures are considered as being isomorphic if $m \to \infty$.

The free energy of the glass can be represented as an expansion in which the first two terms are determined by the $\overline{c}_i$ and $w_{ik}(r_p)$ values:

$$G_0(P, T) = N\left[ \sum_i \overline{c}_{i,g} g_i(P, T) + \sum_{i,k} \overline{c}_{i,g} w_{ik}(r_p) g_{ik}(P, T) \right] \qquad (8)$$

Here we put $w_{ik}(r > r_p) = 0$ assuming that the pair correlations at $r > r_p$ are negligible; $g_{ik}(P, T)$ is the mean free energy contribution from the pair interaction of the associate $i$ with $k$. At operations with the Eq. (8) it has to be taken into account that $\sum_i c_{i,g} = 1$ and $\sum_k w_{ik} = 1$.

In the approximation of Eq. (8) all structural states of the glass with identical $\overline{c}_i$ and $w_{ik}(r_p)$ are taken as isoconfigurational states having the same free energies. The expression for free energy should also allow for the contribution from the intercluster boundaries

$$G(P,T) = G_0(P,T) + G_s(P,T)$$
$$G_s(P,T) = N_s\left(\overline{g}_s(P,T) - \overline{g}_0(P,T)\right)$$

(9)

where $N_s$ is the number of molecules in the cluster boundaries; $\overline{g}_s(P,T)$ and $\overline{g}_0(P,T)$ is the mean free energy per molecule within boundary and bulk respectively. The contributions of other defects into the free energy can be included too, but it is reasonable to separate these contributions explicitly only if we are interested in their related phenomena.

### 2.1.3. Polyamorphous transformations.

The free energy of the glass represented in the form of Eqs. (8) and (9) enables us to construct the phenomenology of polyamorphous transformations. First note that the expression for the free energy includes both the free energies $g_i(P,T)$ and $g_{ik}(P,T)$ describing the thermodynamic properties of associates, as well as the "frozen in" concentrations $\overline{c}_i$ and correlations $w_{ik}(r_p)$. The $n$-th order phase transition entails a jump-like change in the $n$-th order derivative of the free energy with respect to pressure and temperature. If $\overline{c}_i$ and $w_{ik}(r_p)$ are considered to be constant, while $P, T$ are considered as independent variables, a polyamorphous transition is only possible on discontinuity of the derivatives of the functions $g_i(P,T)$ and $g_{ik}(P,T)$. Such a phase transition is *isoconfigurational*. The isoconfigurational transition is caused by the changes in the associates leaving their concentrations and mutual positions unaltered. During the isoconfigurational phase transition the system remains in the same basin but the derivatives of the free energy are not smooth, at least for some types of associates or associate correlations. Isoconfigurational phase transitions are reversible since the topology of associates remains unaltered.

Isoconfigurational polyamorphism is a special case of the general polyamorphism during which both the derivations of $g_i(P,T)$ and $g_{ik}(P,T)$, as well as the magnitudes $\overline{c}_i$ and $w_{ik}(r_p)$, change in a jump-like manner.

Transformability of the glass structure stems from the nature of the "frozen in" configurations. Indeed, the parameters $\overline{c}_i$ and $w_{ik}(r_p)$ correspond to the free energy minimum only at $T > T_g$. This is not truly correct anymore at $T < T_g$ because Eqs. (4) - (5) do not hold in this case. The degree of the glass non-equilibrium is evident in the deviation of the free energy from the minimum

$$\Delta G(P,T) \approx N\sum_i [c_i(P,T) - c_{i,g}]g_i(P,T)$$

(10)

Here $c_i(P,T)$ stands for equilibrium fractions (5). Equation (10) is a simplified form of $\Delta G(P,T)$ since it does not allow for the contribution from pair interactions of the associates. It is clear that the structure relaxation induced by the thermodynamic driving force, Eq. (10), is retarded only by the slow kinetics of structural rearrangements. The slowing down of the thermally activated relaxation processes with a lowering temperature is the reason why the polyamorphous transformation to a glass, manifests itself as a jump-like change of physical quantities and is only observed when the kinetic requirement is met such that

$$\tau_{12} < \tau_{obs}$$

(11)

Here $\tau_{12}$ is the time during which phase 1 transforms into phase 2. The kinetic requirement of Eq. (11) can be fulfilled only if the activation barriers controlling the phase transformation decrease considerably or if the phase nonequilibrium is large enough and there is a mechanism of athermic relaxation similar to martensite transformations (see e.g. [30]).

Since we have to deal with the multiplicity of structural states, it is appropriate to specify the meaning of $\tau_{12}$. We can denote the sets of isoconfigurational states of two compositionally identical glass phases as

$$\{S_1\} = (s_1^{(1)}, s_1^{(2)}, \ldots s_1^{(W_1)})$$
$$\{S_2\} = (s_2^{(1)}, s_2^{(2)}, \ldots s_2^{(W_2)})$$

(12)

The numbers of structural states of the phases, $W_1$ and $W_2$, are determined by Eq. (2)

Each of the structural states $s_i^{(l)}$, $i$=1,2 has a free energy

$$G_{1,2}(P,T) = N\mu_{1,2}(P,T)$$

(13)

$$\mu_1(P,T) = \sum_i c_{1,i}[g_{1,i}(P,T) + \sum_k w_{1,ik} g_{1,ik}(P,T)]$$
$$\mu_2(P,T) = \sum_i c_{2,i}[g_{2,i}(P,T) + \sum_k w_{2,ik} g_{2,ik}(P,T)]$$

(14)

which is dependent only on the upper index indicating the number of the structure state.

Assume that the glass transition leads to the formation of the state $s_1^{(l)}$ with the chemical potential $\mu_1(P,T)$. After crossing the phase co-existence curve given by the equation

$$\mu_1(P,T) = \mu_2(P,T)$$

(15)

the states of phase 2 have lower free energy than those of phase 1. As a result, a polyamorphous phase transformation is advantageous thermodynamically and conceptually the state $s_1^{(l)}$ can transform into any of the states $\{S_2\}$. It is clear that in the approximation of Eq. (8), the thermodynamic driving force of the transformation $\Delta\mu = \mu_1(P,T) - \mu_2(P,T)$ is independent of which of the states of $\{S_2\}$ is realized by the transition. However, the time of transformation is essentially dependent on the microscopic structure of the final state because the transition causes rearrangement of the SRO. We take $\tau_{lm}$ as the characteristic time of the $s_1^{(l)} \to s_2^{(m)}$ transformation. Among a large body of $\tau_{lm}$ values, there is the smallest one

$$\tau_l^* = \min_m \tau_{lm}$$

(16)

The condition in Eq. (16) determines which of the states $\{S_2\}$ will result from the transformation of $s_1^{(l)}$. The resulting state is numbered $m(l)$. $\tau_l^*$ averaged over all states $\{S_1\}$ determines the characteristic time of the phase transformation in Eq. (11)

$$\tau_{lm} \quad \tau_{12} = <\tau_l^*>$$

(17)

It is obvious that the time is the shorter if the fewer structural rearrangements are required for the polyamorphous transformation and the lower are the barriers of these rearrangements. The proportions of structural changes during the $s_1^{(l)} \to s_2^{(m)}$ transformation are characterized by $c_{i}$, g variations,

$$\Delta c_i = c_{1,i}^{(l)} - c_{2,i}^{(m)}$$

(18)

and by the overlapping parameter

$$q_{lm} = \sum_i c_{1,i}^{(l)} \left\langle c_{1,i}^{(l)}(x) c_{2,i}^{(m)}(x+r_p) \right\rangle$$

(19)

The quantity $\Delta c = (\Delta c_1, \Delta c_2, \ldots, \Delta c_{n_s})$ is the vector of $n_s$ components in the space of $\Delta c_i$-components. Choosing the space metric in the form

$$|\Delta c| = \frac{1}{2}\sum_i |\Delta c_i|$$

(20)

we can estimate the extent of the difference between the two structures. The metric in Eq. (20) is suitable because it includes the relation $0 \leq |\Delta c| \leq 1$ similar to the relation $0 \leq c_i \leq 1$, that applies for the fraction $c_i$ of orientational associate type $i$.

The overlapping parameter $q_{lm}$ describes changes of the pair correlators. During the isoconfigurational transformation, when $|\Delta c| = 0$, the overlapping parameter $q_{lm}$ is equal to unity. In contrary, it is a small quantity if the structure of the state $s_2^{(m)}$ is essentially different from that of $s_1^{(l)}$.

The $\tau_l^*$-value is expected to be independent $l$ in the thermodynamic limit. This means that each of the states $s_1^{(l)}$ has a $|\Delta c|$- spaced neighboring state $s_2^{(m)}$ with the overlapping parameter $q_{lm}$ and at that the magnitudes $|\Delta c|$ and $q_{lm}$ are independent of the index $l$.

## 2.2. Low temperature polyamorphous transformations in orientational glasses based on doped fullerite $C_{60}$.

Using the results of the previous section, we can now analyze the polyamorphous transformations revealed experimentally in the orientational glasses based on doped fullerite $C_{60}$. At low temperatures the free energy of the molecule with the $i$-th orientational SRO (or associate type) can be written as (disregarding the associate interaction):

$$g_i(P,T) = u_{0i} + \sum_{l=1}^{n_\Omega} \frac{1}{2} \left[ \hbar \Omega_i^{(l)} - T \exp(-\hbar \Omega_i^{(l)} \beta) \right] + g_{ph}(P,T) + Pv \qquad (21)$$

Here $u_{0i}$ is the energy of the molecule disregarding the zero point vibrations, whose contributions are described by the first terms under the summation sign in Eq. (21), $\Omega_i^{(l)}$ are the frequencies of libration and intramolecular vibrations; $n_\Omega$ is the number of their frequencies, $g_{ph}(P,T)$ is the phonon contribution presumably independent of the type of orientational SRO and minor changes in the lattice parameter during phase transformation; $v$ is the specific volume. Of all the terms with $\sim \exp(-\hbar \Omega_i^{(l)} \beta)$, only those with frequencies close to the lowest $\Omega_i^{(l)}$ can contribute appreciably. The number of significant terms $\sim \exp(-\hbar \Omega_i^{(l)} \beta)$ can be denoted as $n_\Omega$. Since the frequencies of intramolecular vibrations are much higher than the libration frequencies and the number of the latter frequencies is 3, we obtain $n_\Omega = 3$.

Assume that $\overline{\Omega}_i$ is the mean value of the frequencies for the $i$-th orientational SRO, and $(\delta\Omega)_i^2$ is variance of these frequencies. We consider the low temperature region where $T << \hbar \overline{\Omega}_i$. At the same time we assume that $(\hbar \delta\Omega_i / 4T)^2 << 1$. Taking into account only the second order corrections in parameter $(\hbar \delta\Omega_i / T)$, Eq. (20) can be rewritten as

$$g_i(P,T) = u_{0i}(P,T) - Tn_\Omega \exp(-\hbar\overline{\Omega}_i \beta)[1 + (\hbar\delta\Omega_i / 4T)^2] + g_{ph},(P,T) + Pv \qquad (22)$$

Substituting the expression for $g_i(P,T)$ into Eq. (8) and neglecting the contribution of the associate interaction, we can obtain with the most minimal loss in accuracy

$$G(P,T) = N\{\overline{\varepsilon}(P,T) - Tn_\Omega \exp(-\hbar\overline{\Omega}\beta)[1 + (\hbar\delta\Omega / T)^2] + g_{ph} + Pv\} \qquad (23)$$

where

$$\overline{\varepsilon}(P,T) = \sum_i c_i u_{0i}(P,T) \qquad (24)$$

$$\exp\left(-\hbar\overline{\Omega}\beta\right) = \sum_l c_i \exp\left(-\hbar\overline{\Omega}_i\beta\right), \quad \delta\Omega^2 = \sum_i c_i(\overline{\Omega}_i - \overline{\Omega})^2 \qquad (25)$$

Eqs. (22) – (25) show that the approximation chosen is an effective medium approximation. Inclusion of the associate interactions will naturally lead to renormalization of $\overline{\varepsilon}$ and $\overline{\Omega}$.

The above expression suggests that the phase transition is possible if there are two minima of free energy, Eq. (23), as function of $\{c_i\}$. We use $\mu_1(P,T)$ and $\mu_2(P,T)$ to denote the chemical potentials of the phases (as previously used in Eq. 14), and given by the braced terms in Eq. (23), at minima 1 and 2, respectively. The transition temperature $T_e$ is then determined by the Gibbs equation

$$\mu_1(P,T) = \mu_2(P,T) \tag{26}$$

which after cancelling of the phonon contributions, disregarding the term $P(v_1 - v_2)$ and the corrections quadratic in $\hbar\delta\Omega\beta$, gives:-

$$\varepsilon_1(P,T) - T_e n_\Omega \exp(-\hbar\overline{\Omega}_1\beta) = \varepsilon_2(P,T) - T_e n_\Omega \exp(-\hbar\overline{\Omega}_2\beta) \tag{27}$$

To get solution Eq.(27) one should note that the temperature dependence of energies $\overline{\varepsilon}_1, \overline{\varepsilon}_2$ and frequencies $\overline{\Omega}_1, \overline{\Omega}_2$ can be neglected at low temperatures where $\overline{\varepsilon}_{1,2}(P,T) \approx \overline{\varepsilon}_{1,2}(P,0) \equiv \overline{\varepsilon}_{1,2}$ and $\overline{\Omega}_{1,2}(P,T) \approx \overline{\Omega}_{1,2}(P,0) \equiv \overline{\Omega}_{1,2}$. As result we have

$$T_e n_\Omega [\exp(-\hbar\overline{\Omega}_2\beta) - \exp(-\hbar\overline{\Omega}_1\beta)] \approx n_\Omega \hbar(\overline{\Omega}_2 - \overline{\Omega}_1)e^{-\hbar(\overline{\Omega}_1+\overline{\Omega}_2)\beta/2} = \overline{\varepsilon}_2 - \overline{\varepsilon}_1$$

or

$$T_e = -\frac{\hbar(\overline{\Omega}_1 + \overline{\Omega}_2)}{2}\left\{\ln\left[\frac{\overline{\varepsilon}_2 - \overline{\varepsilon}_1}{n_\Omega \hbar(\overline{\Omega}_2 - \overline{\Omega}_1)}\right]\right\}^{-1} =$$

$$= -\frac{\hbar(\overline{\Omega}_1 + \overline{\Omega}_2)}{2}\left\{\ln\left[\frac{\overline{\varepsilon}_2 - \overline{\varepsilon}_1}{3\hbar(\overline{\Omega}_2 - \overline{\Omega}_1)}\right]\right\}^{-1} \tag{28}$$

## 2.3. Thermal expansion coefficient.

In the previous section our attention was focused on the thermodynamics of a low temperature transition. The contribution from the cluster boundaries was neglected. Indeed, this contribution is negligible at high (near $T_g$) temperature. However, at low temperatures when the phonon contribution is small (see the Introduction), the tunneling states situated at the cluster boundaries start to play a significant role.

An important feature of the low temperature excitations of the tunneling states is a weak temperature dependence of the relaxation time. Owing to this feature, the system of tunneling states (TS) can be considered thermodynamically as an equilibrium subsystem of the glass. On polyamorphous transformation, the cluster boundary structure, the number of tunneling states and the density of their low–energy states all change. These changes come as evidence of the polyamorphous transition at low temperatures when it is possible to separate the TS contribution to the temperature coefficients. In this section we analyze the component $\alpha_{TS}(T)$ of the thermal expansion coefficient $\alpha(T)$. Existence of polyamorphism in orientational glasses based on doped $C_{60}$ was detected from the hysteresis of $\alpha(T)$.

Let us consider an isolated TS. The free energy of the TS with the level splitting $E$ can be written as

$$g_{TS}(T,v) = \varepsilon_0(v) - T \ln 2 \cosh(E(v)\beta/2) + \varepsilon_{el}(v) \tag{29}$$

Here $\varepsilon_0(v)$ is the sum of energies of the well bottom and zero oscillations; $v$ is the TS volume, $\varepsilon_{el}(v)$ is the elastic deformation energy.

$$E^2(v) = \Delta^2 + \Delta_0^2(v)$$
$$\Delta_0 = \hbar\Omega e^{-\lambda}, \quad \lambda \sim (2IV_0/\hbar)^{1/2}\varphi \tag{30}$$

where $\Omega$ is the libration frequency, $I$ is the inertia moment, $V_0$ is the height of the hump separating two potential well minima, $\phi$ is the angle of rotation to get from one minimum to the other, $\Delta$ is the difference between the minima depths.

The two – well potential at $\Delta = 0$ can be represented as a sine curve within the interval, $-\varphi < \theta < \varphi$

$$U(\theta) = U_0(v)[\cos(2\pi\theta/\phi - 1] + Const \tag{31}$$

and supplemented with infinitely high walls at $\theta = \pm\Delta\theta$. In this case

$$\varepsilon_0(v) = -2U_0(v) + \hbar\Omega_0/2 + Const \tag{32}$$

and the hump height is $V_0 = 2U_0(v) - \hbar\Omega_0/2$. Here $\Omega_0$ is the zero oscillation frequency.

The origin of the elastic energy $\varepsilon_{el}(v)$ is as follows. If the tunneling barrier is very high and $\Delta_0 \to 0$, the TS gains the volume $v_0$ which can be found from the condition of equilibrium between the TS and the surrounding molecules. At finite $\Delta_0$ – values, the correction $\Delta v_{TS}$ appears. As a result,

$$v = v_0 + \Delta v_{TS} \tag{33}$$

(The necessity of including this correction in the TS contribution to the thermal expansion coefficient was first considered in [31]. Using special assumptions, $\Delta v_{TS}$ for methane was calculated in [32].)

When the addition $\Delta v_{TS}$ appears, the tunneling state becomes a dilatation center whose elastic energy is [33]

$$\varepsilon_{el}(v) = \frac{1}{2}f_{el}v_0\left(\frac{\Delta v_{TS}}{v_0}\right)^2 \tag{34}$$

where

$$f_{el} = \frac{4}{9}\mu\frac{1+\nu}{1-\nu} \tag{35}$$

Here $\mu$ is the shear modulus and $\nu$ is the Poisson coefficient. Now $\Delta v_{TS}$ can be found by minimizing the free energy of Eq. (29) in $v$. From the requirement of the minimum

$$\frac{\partial g_{TS}(v)}{\partial v} = 0 \tag{36}$$

we have

$$2\left[1 - \frac{1}{4}\tanh\frac{E\beta}{2}\cdot\frac{\lambda\Delta_0^2}{EV_0}\right]\frac{\partial U_0}{\partial v} = f_{el}\frac{\Delta v_{TS}}{v_0} \tag{37}$$

Hence,

$$\frac{\Delta v_{TS}}{v_0} = 2f_{el}^{-1}\left[1 - \frac{1}{4}\frac{\lambda\Delta_0^2}{EV_0}\tanh\frac{E\beta}{2}\right]\frac{\partial U_0}{\partial v} \tag{38}$$

We may state that the derivative $\partial U_0(v)/\partial v$ is negative since the well depths in Eq. (30) and the height of the hump between the wells should decrease when the volume increases. Note that Eqs. (37) and (38) were derived neglecting the dependence of $\Delta$ on the volume, because we assumed that the $E(v)$ variation was mainly caused by the changes in the depths of the wells and the separating barrier (see Eqs. (31), (32)).

Taking into account that the second term in bracket of Eq. (38) is always less than unity (since $\Delta_0 \leq E < U_0$ and the $\tanh(E\beta/2) \leq 1$), we can find that $\Delta\nu_{TS} < 0$, i.e. the tunneling states compress the sample. Since $\tanh(E\beta/2)$ decreases with rising temperature, then

$$\alpha_{TS}(E, \Delta_0, T) = \frac{1}{\nu_0}\frac{\partial \Delta\nu_{TS}}{\partial T} = \frac{1}{4T^2}\frac{\lambda\Delta_0^2}{f_{el}V_0}\frac{1}{\coth^2(E\beta/2)}\frac{\partial U_0}{\partial \nu} < 0 \qquad (39)$$

To estimate the total contribution of the tunneling states to the thermal expansion coefficient, Eq. (39) should be integrated with respect to all the TS. Using $f_{TS}(E, \Delta_0)$ for the distribution function of TS as a function of both $E$ and $\Delta_0$, and denoting the TS concentration (the number of TS per molecule) as $c_{TS}$, we obtain:

$$\alpha_{TS}(T) = c_{TS}\int\limits_0^\infty\int\limits_0^\infty f_{TS}(E, \Delta_0)\alpha_{TS}(E, \Delta_0, T)dEd\Delta_0 \qquad (40)$$

The distribution function $f_{TS}(E, \Delta_0)$ is assumed to be normalized to unity.

The low temperature ($T \leq 1$K) contribution of TS to the thermodynamic coefficients is usually described using a function homogeneous over the finite interval of $\Delta$ - and $\lambda$ - values. In terms of the variables $E$ and $\Delta_0$ it is

$$f_{TS}(E, \Delta_0) \sim \frac{\overline{P}E}{\Delta_0\sqrt{E^2 - \Delta_0^2}} \qquad (41)$$

Here $\overline{P}$ is a constant. We are interested in how the TS system influences the thermal expansion coefficient in the temperature interval of about 20K, i.e. far from the region where the low temperature anomalies are observable. In this case the distribution function of Eq. (41) can be invalid for a reasonable interpretation of experimental results.

The phonon contribution into the thermal expansion coefficient can be found using the phenomenological expression

$$\alpha_{ph}(T) = \Gamma C_{ph}(T), \qquad (42)$$

where $\Gamma$ is the Gruneisen coefficient and $C_{ph}(T)$ is the phonon contribution to heat capacity. It is known that at low temperatures $C_{ph}(T) \sim T^3$. As a result, $\alpha_{ph}(T)$ becomes comparatively low. The total thermal expansion coefficient is

$$\alpha(T) = \alpha_{ph}(T) + \alpha_{TS}(T), \qquad (43)$$

where $\alpha_{TS}(T)$ is determined by Eqs. (39) and (40), and can become negative in the low temperature region, where $\alpha_{ph}(T) < \alpha_{TS}(T)$.

## 2.4. Relaxation process.

The dilatometric technique applied to orientational $C_{60}$–based glasses permits investigation of the relaxation processes whose characteristic times are no longer than $\tau_{obs} \sim 10^3$ s. Three important relaxation processes can be distinguished in the studied orientational glasses. One of them is the polyamorphous transformation time $\tau_{12}$ included in the kinetic criterion of Eq. (10). The other two are the times during which the distribution functions of elementary excitations in the phonon-libron system and in the TS system come to equilibrium. We start with the polyamorphous transformation.

The phase transformation during the first–order phase transitions can be described within the framework of the Avrami–Kolmogorov model [34]. According to this model, after crossing the phase co-existence curve, the fraction of the new phase changes by the law:

$$V_2(t) = V_0[1 - \exp(-At^n)] \qquad (44)$$

Here $V_0$ is the total volume, $A$ is a constant dependent on the rates of the new phase nucleation and growth, $n$ is the Kolmogorov exponent and dependent on the system dimensions. For example, with uniform nucleation at the rate $I$ and the growth constant $u_g$ in the 3-dimensional space $A = I u_g^3$ and $n=4$.

If nucleation is non-uniform occurring mostly in special regions (e.g. at the grain boundaries) and the nucleation rate is high, the time of the phase transformation is determined by the growth rate of the new phase precipitates. In this case we can put in Eq. (44)

$$A \approx B u_g, \quad n = 1, \tag{45}$$

where $B$ is the density of the sites where the new phase nucleates. If nucleation occurs at the grain boundaries, $B \approx 2/l$, where $l$ is the average linear grain size.

To identify the scenario of phase transformation, we estimate the bulk thermodynamic driving "force" controlling the rate of the transformation. Under a constant low pressure in the first order of $T - T_e$, we have

$$\Delta \mu_{12}(T) = \mu_1(T) - \mu_2(T) = -\Delta \varepsilon_{12}(T_e)(T - T_e)/T_e;$$
$$\Delta \varepsilon_{12} = \varepsilon_1 - \varepsilon_2 \tag{46}$$

Since in the general case the phase transformation entails changes in the volume and in SRO, we can separate these contributions to $\Delta \varepsilon_{12}$:

$$\Delta \varepsilon_{12} = \Delta \varepsilon_{el} + \Delta \varepsilon_{or} + \Delta \varepsilon_{intra} \tag{47}$$

where

$$\Delta \varepsilon_{el} = \frac{K}{2} \left( \frac{\Delta v}{v} \right)^2 v \tag{48}$$

$K$ is the bulk elastic modulus, $v$ is the specific volume. $\Delta \varepsilon_{el}$ shows what amount of energy is stored in the sample due to the volume variation in the transformation process. $\Delta \varepsilon_{or}$ accounts the energy changes due to the orientational order difference and $\Delta \varepsilon_{intra}$ allows for the changes in the intramolecular structural state.

Since the free energy and entropy decrease during phase transformation, the energy per molecule in the low temperature phase (phase 1) is lower than that in the high temperature phase. Thus, $\Delta \varepsilon_{12} < 0$. Since $\Delta \varepsilon_{el}$ is always positive, the sum $\Delta \varepsilon_{or} + \Delta \varepsilon_{intra}$ should be negative and its absolute value exceeds $\Delta \varepsilon_{el}$.

It is beyond reason to expect that the intramolecular ground state can change as $T \to 0$. Otherwise, the phase transition could occur in pure $C_{60}$ since the mutual positions and orientations of molecules (essential for molecular interactions) are less important. Therefore, in Eq. (47) we can put $\Delta \varepsilon_{int} = 0$.

It is quite simple to estimate $\Delta \varepsilon_{el}$. Since in fullerites $K \approx 10.3 \, GPa$, $v \sim \cdot 10^3 \, \text{Å}^3$ and, as it is seen from our data, $\Delta v / v \sim 10^{-3}$, we have $\Delta \varepsilon_{el} \sim 10^{-1} \, K$.

$\Delta \varepsilon_{or}$ is related to the changes in SRO. According to Eq. (24),

$$\Delta \varepsilon_{or} = \sum \Delta c_i \varepsilon_i \tag{49}$$

Taking into account that $\Delta c_i < 1$ (e.g., $\Delta c_i \sim 10^{-1}$) and the energy difference between two neighbouring molecules can amount to $\sim 10^2 \, K$ due to the change of the mutual orientation [1, 2], we can put $\Delta \varepsilon_{or} \sim 10 \, K$.

The above estimates suggest that no isoconfigurational transition (for such transition $\Delta \varepsilon_{or} = 0$) occurs in the investigated orientational glasses. In our case the phase transformation is a cooperative change of the SRO and the lattice parameter. Since the phase transformation is accompanied by a change in the volume, it is most likely that the transformation process starts at the sample surface or at the grain boundaries. In this case elastic relaxation is possible where the new phase nucleates. The ensuing quasi–one–dimensional growth of the nuclei changes the lattice parameter at the phase interface. In this scenario of the phase transformation, the SRO is disturbed at the phase interface and this suppresses the potential barriers for orientational rearrangements. It is therefore quite easy to overcome the barriers under the action of local inner elastic stresses and weak thermal fluctuations. The diffusion–free mechanism of relaxation of elastic stresses at the transformation front caused by an inconsistency between the lattice parameters deserves special attention and will be considered elsewhere. Here we can conclude that the phase transition in the studied orientational glasses can be described by Eqs. (44), (45), the structural transformation is a nondiffusive process and the transformation rate is mainly determined by $\Delta\varepsilon_{or}$. As seen in Eq. (45), the characteristic time of phase transformation is

$$\tau_{12} \sim l / u_g \qquad (50)$$

If the above estimate $\Delta\varepsilon_{or} \sim 10K$ is correct, then a considerable amount of heat is released at the front of the phase 2 $\rightarrow$ phase 1 transformation at $|T - T_e| \sim T_e$. This causes local heating and accelerates the transformation process. On a reverse transformation, above $T_e$, the heat is absorbed at the transformation front, which hampers a thermoactivated overcoming of even low potential barriers. We should expect slowing – down of the reverse transformation in this case.

The relaxation time of the phonon–libron system, $\tau_{ph}$, is determined by the free path of the phonons that are scattered at these excitations and are responsible for their thermalization and the equilibration of the distribution function. This is obviously the shortest of all the relaxation times under consideration, $\tau_{ph} << \tau_{12}, \tau_{TS}$. This relaxation process therefore produces very little effect on the polyamorphous transformations of our interest.

The relaxation time of the TS system is mainly determined by the tunneling time [10],

$$\tau_{TS}^{-1} = A\Delta_0^2 E \coth(E\beta) \qquad (51)$$

Here $A$ - is a certain constant. At low temperatures, when $E\beta \geq 1$, the last factor in the right–hand part of the equation has a value of the order of unity. As a result, $\tau_{TS}$ is only slightly dependent on temperature in this case.

### 3. Comparison with experiment.

Worth noting is noted that the phase transition is revealed in fullerites doped by gases while in pure fullerite this phenomenon was not observed. The dopants change molecular interactions and the potential energy landscape. As result structural, thermodynamic and kinetic properties are changed. For this reason the thermodynamic quantities and kinetic coefficients are depending on the dopant concentrations. The put forward theoretical model is valid inspite these dependencies are not known.

It was hardly possible in this study to find experimental temperatures $T_e$ of coexistence of the glass phases (phase transformation temperature) in the investigated materials. This is because phase transformations are inevitably smeared. $T_e$ is expected to be somewhere within the region of the thermal expansion hysteresis.

In our experiments the upper boundary of the hysteresis for Xe-$C_{60}$ was not reached. It seems plausible that the hysteresis and $T_e$ depend on the type of the doping gas. The X-ray diffraction measurement [35] of the lattice parameter $a$ of Xe-$C_{60}$ in a wide interval above 7K also revealed a hysteresis of $\alpha(T)$ in the region 7-65K. The largest width of the hysteresis loop was observed at $T \approx 20K$. Like in our experiments, in [35] $\alpha(T)$ was lower on cooling from high temperature and higher on heating from 7K. This counts in favour of polyamorphism, which we detected in Xe-C$_{60}$. Besides, it shows that $T_e$ of phase coexistence is above 7K, being presumably $T_e \approx 20K$.

When using Eq. (28) to estimate $T_e$, we should remember that according to [36, 37], the characteristic frequency of libron vibrations is about $\hbar\Omega_1 \approx \hbar\Omega_2 \approx 40K$. Taking $\varepsilon_1 - \varepsilon_2 \sim 10K$ and $|\Omega_1 - \Omega_2| / \Omega_1 \sim 10^{-4} - 10^{-3}$ (which seems quite reasonable), $T_e \sim 10K$ (according to Eq. (28)). This is a rough estimate, but since the difference quotient $(\varepsilon_1 - \varepsilon_2)/(\Omega_1 - \Omega_2)$ appears in Eq. (28) in the logarithmic form, $T_e$ has only a weak dependence on the choice of these quantities.

We were able to separate the positive ($A$) and negative ($B$) contributions to the thermal expansion coefficients owing to their considerably different relaxation times under jump-like changes of temperature.

The positive contribution $A(T)$ to the $\alpha(T)$ made by the low-frequency lattice excitations (phonons and librons) and its relaxation time $\tau$ ($T$) are little dependent on the type of the doping gas. $A(T)$ and $\tau$ ($T$) are practically insensitive to polyamorphous transformations.

The negative contribution to the heat capacity $B(T)$ is in our opinion connected with the TS system which undergoes significant restructuring during the polyamorphous transformation. $B(T)$ and its relaxation time are strongly dependent on the type of doping gas. The negative contribution to the $\alpha(T)$ of C$_{60}$ doped with gases is described below.

Taking

$$B(T)=\alpha_{TS}(T) \qquad (52)$$

and assuming that $\alpha_{TS}$ is described by Eqs. (38), (39), we can understand from the analysis of experimental results how the TS system changes during the polyamorphous transformation. First note that the temperature dependence of the contribution to the $\alpha(T)$ from the TS system on splitting of the energy levels $E$, $\alpha_{TS}(E,\Delta_0,T)$, has a maximum at $T \approx E/2$. This function decreases quite rapidly with lowering temperature and slowly, $\sim T^{-2}$, at high temperatures. A linear superposition of these functions, Eq.(40), tends rapidly to zero at $T < E_{\min}/2$ and decreases as $\sim T^{-2}$ at high temperatures $T > E_{\max}/2$. $E_{\min}$ and $E_{\max}$ are the lowest and highest values of $E$.

Eq. (40) holds for an equilibrium stationary TS system and can be used when its relaxation time $\tau_{TS}$ is much shorter than the time of observation. In Eq. (50) $\tau_{TS}$ is inversely proportional to squared $\Delta_0$. As a result, two-level systems with $\tau_{TS}\overline{\Omega\Delta}_0 < 1$ do not contribute to thermodynamic quantities, in particular, to $\alpha(T)$'s. Thus, the coefficient $c_{TS}$ in Eq. (40) has to include only the concentration of "active" TS coming to equilibrium during the time of observation. Therefore, the negative contribution $B(T)$ appearing during the polyamorphous transformation at $T < T_e$ should be attributed to a change in the concentration of active TS undergoing the levels splitting $E$ comparable with the temperature.

The structural transformations during the reverse orientational transition at $T>T_e$ are very slow near $T_e$. Therefore, the parameter $c_{TS}$ is assumed to be temperature – independent at $T\approx T_e$, but away from $T_e$ it depends both on time and on thermal prehistory, which is important for the analysis of experimental data.

We lack information to take into account the polyamorphous transformation in the relaxation kinetics of the TS system, but Eqs. (38) and (39) seem to be good for an adequate analysis of the experimental results in a wide temperature interval near $T_e$.

$B(T)$ measured on Xe-C$_{60}$ and D$_2$-C$_{60}$ are shown in Fig. 13(a). These values correspond to the low temperature phase having an increased specific volume. In both the samples, the $B(T)$ behaviour has characteristic features: (i) the presence of maxima and (ii) a sharp decrease in the low temperature region below the lower maximum.     Besides, the $B(T)$ peak at $T$=6  in D$_2$-C$_{60}$, it is narrower than that following from Eq. (38) as shown in Fig. 13(b).

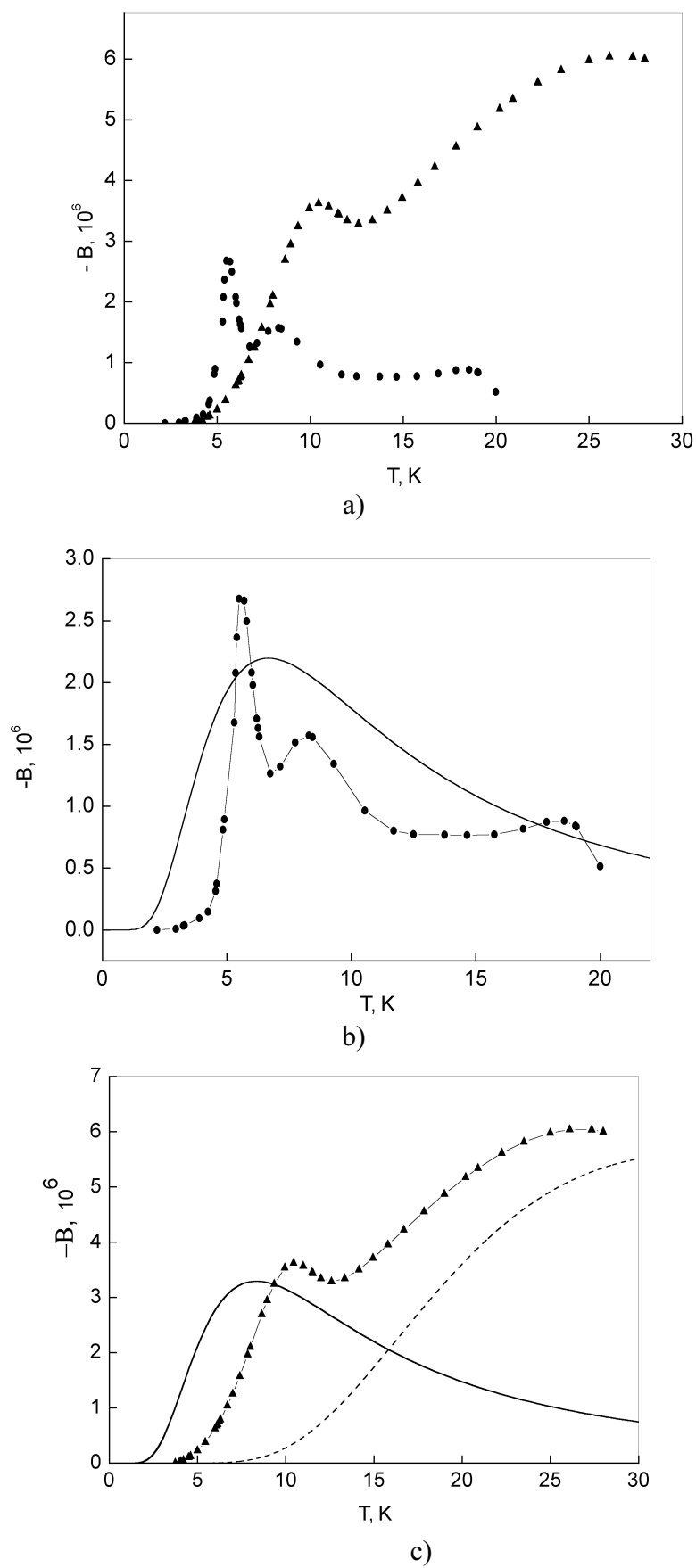

Fig. 13. Temperature dependence of $B(T)$.
   a) Experimental data for $D_2 - C_{60}$ (●) and Xe-$C_{60}$ (▲) ;
   b) Fitting by Eq. (53) (solid line) for experimental data $D_2$-$C_{60}$ (●);
   c) Fitting by Eq. (54) (dashed line) for experimental data Xe-$C_{60}$ (▲)
      and fitting by Eq. (53) (solid line) for experimental data $D_2$-$C_{60}$.

The temperature dependence of the negative contribution can be analysed in terms of Eqs. (38), (39) and (52). We can put $f(E,\Delta)=f_1\delta(E-E_1)$ because with an equilibrium stationary density of TS, a distinct $\alpha_{TS}(T)$ peak is possible only for a narrow distribution of $f(E, \Delta_0)$ concentrated near a certain value $E_1$. In this case

$$\alpha_{TS}(T) = c_{TS}(E_1)\alpha(E_1, \Delta_0, T) \qquad (53)$$

The $B(T)$ values fitted by Eq. (53) for the $D_2$-$C_{60}$ sample are shown in Fig. 13 b. It is seen that Eq. (53), in which $c_{TS}$ comes as independent of temperature and time, offer a rather rough description of experimental data. If $T_e$ is close to the peak, $T_1 \approx E_1/2$ and the fast decrease in $B(T)$ at $T>T_1$ can be attributed to the reverse polyamorphous transformation. This explanation seems quite reasonable regarding that the time of the reverse transformation at $T$=10.5K is about $5\bullet10^3$ (Fig. 12), i.e. comparable to the time of measuring $B(T)$. But the experimental results available are not sufficient for a more detailed analysis.

The more dramatic (than the assignment of Eq. (53)) decrease in $B(T)$ at $T<T_1$ suggests that the distribution of TS levels is temperature–dependent at $T<T_e$ as well. In the case of an equilibrium TS system in this $T$-interval (like the low temperature phase of an orientational glass) this behavior can be explained through a deeper insight into the thermodynamics of the TS system. If the TS density is non-equilibrium and varies with time during measurement, it is necessary to analyze the relaxation kinetics of the system, which is beyond the scope of this study.

The data on the kinetics of the reverse polyamorphous transformation (Fig. 12) show that at least the long–time asymptotic of Eq. (43) can be described by the simple exponent $\sim[1-\exp(t/\tau)]$, i.e. the Kolmogorov exponent is equal to unity at least at the late stage of the phase transformation (see Eq. (44)). This means that at this stage the polyamorphous transformation proceeds as a one-dimensional growth of the stable phase at a two- dimensional interphase boundary with the metastable (nonequilibrium) phase.

The dependence $B(T)$ for $C_{60}$ doped with Xe should be analyzed taking into account the two peaks with a slight dip in between that are observed in this case. Following the reasoning used for the $D_2$-$C_{60}$ system, we arrive at the conclusion that in Xe-$C_{60}$ the TS density is concentrated near $E_1\approx10$K and $E_2\approx40$K. Fig. 13c shows the experimental results fitted by the expression

$$\alpha_{TS}(T) = c_{TS}(E_1)\alpha_{TS}(E_1, \Delta_0, T) + c_{TS}(E_2)\alpha_{TS}(E_2, \Delta_0, T) \qquad (54)$$

where $c_{TS}$ $(E_1)$ and $c_{TS}$ $(E_2)$ are the fitting constants, $\varepsilon_{TS}(E,T)$ is described by Eq. (38). The data obtained are in good qualitative and quantitative agreement. The better fitting in this case (compared to that for $D_2$-$C_{60}$) suggest that in Xe-$C_{60}$ the TS density is weakly dependent on temperature in a wide interval. This conclusion is supported by the measurement in [35] where it was shown that on heating Xe-$C_{60}$ the reverse polyamorphous transformation was complete only at $T \approx 60$K . The process is rather slow and cannot manifest itself in the temperature interval of our measurements, at $T<$28K.

However, at low temperatures the coefficient $B(T)$ decreases faster than it is postulated by Eq. (54) containing the best fitting parameters. Like in the case of $D_2$-$C_{60}$, this behaviour is not yet clear.

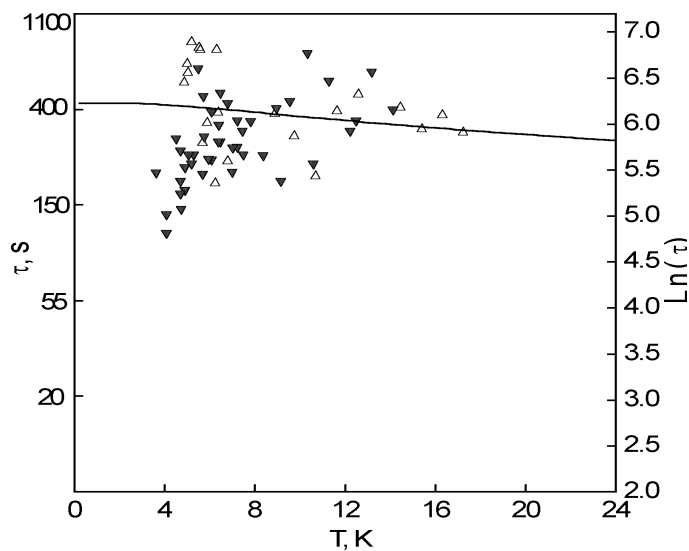

Fig.14. Characteristic time $\tau$ $(T)$ of the negative contribution of the thermal expansion of $H_2$-$C_{60}$ and $D_2$-$C_{60}$.

Experiment:

▼ - $H_2$-$C_{60}$

△ - $D_2$-$C_{60}$

— - description by Eq.(51).

The relaxation times $\tau$ $(T)$ fitted by Eq. (51) for the $H_2$-$C_{60}$ and $D_2$-$C_{60}$ systems are shown in Fig.14. The scatter in the experimental data is too large to expect satisfactory agreement. But it should be noted that the TS system can not be considered as independent on temperature and time as Eq. (51) requires.

### Conclusions.

The dilatometric investigation in temperature range 2-28K shows that the first-order polyamorphous transition occurs in orientational glasses based on $C_{60}$ doped with $H_2$, $D_2$ and Xe. A polyamorphous transition was also detected in our earlier [15] study on $C_{60}$ doped with Kr and He. The hysteresis of thermal expansion caused by the polyamorphous transition (and, hence, the transition temperature) is essentially dependent on the type of the doping gas.

Both positive and negative contributions to thermal expansion were observed in the low temperature phase of the glasses. Within the investigated temperature interval the relaxation time of the negative contribution is considerably longer than that of the positive contribution. This fact has permitted us to separate and analyze both the contributions to the thermal expansion. The positive contribution is found to be due to the low-frequency excitations of the lattice (phonon and librons) and its value is weakly dependent on the type of the doping gas. Arguments are advanced that the negative contribution is due to the tunnel reorientations of some of the $C_{60}$ molecules. The relaxation time of the negative contribution is strongly dependent on the type of the doping gas. This means that the molecules (atoms) of the doping gas affect appreciably the penetrability of the potential barriers separating different orientations of the tunneling $C_{60}$ molecules.

In the high-temperature glassy phase, only the positive contribution to thermal expansion is observed. Its value and relaxation time coincide, within the experimental accuracy, with the corresponding values for the positive contribution in the low temperature phase. This means that the polyamorphous transition discussed here is accompanied first of all by rearrangement of the tunnel state system.

A theoretical model is proposed to interpret these phenomena. The order of magnitude of the polyamorphous transition temperature has been estimated. The estimate agrees with the experimental results. The characteristic time of the phase transformation from the low-$T$ phase to the high-$T$ phase has been found for the $C_{60}$-$H_2$ system at 12K. Its value is an order of magnitude higher than the characteristic time of the tunnel reorientation of $C_{60}$ molecules.

The late stage of the polyamorphous transformation is described well by the Kolmogorov law with the exponent $n=1$. This means that at this stage of tranformation the two-dimensional phase boundary moves along the normal, and the nucleation of the new phase is of no importance.

The arguments are obtained in favor of the assumption that the $H_2$ and $D_2$ molecules dissolved in $C_{60}$ occupy both the octahedral and the smaller tetrahedral interstitial cavities in the $C_{60}$ lattice.

This study was supported by the Science and Technology Center of Ukraine (STCU, project Uzb-116(j)) and National Academy of Sciences of Ukraine, (complex program of fundamental researches "Nanosystems, nanomaterials and nanotechnologies").

We would like to thank Profs. V.D. Natsik, M.A. Strzhemechnyi and A.I. Prokhvatilov for valuable and critical discussion.

**Addition on correcting the text.** After this paper had been sent for publication, we came to know the results on low temperature microhardness of Xe-intercalated fullerite $C_{60}$ [38] which pointed to a considerable increase in the microhardness of $C_{60}$ on its intercalation with Xe. This result agrees with the increase of brittleness of $C_{60}$ when Xe is solved in it.